\documentclass[9pt,twocolumn,twoside]{opticajnl}
\journal{opticajournal} % use for journal or Optica Open submissions

\setboolean{shortarticle}{true}
\usepackage{lineno}
\title{Computer-generated holography enables high-uniformity, high-efficiency depth-of-focus extension in endoscopic OCT}

\author[1,2]{Chengfu Gu}
\author[1,2]{Haoran Zhang}
\author[1]{Qi Lan}
\author[1]{Weiyi Zhang}
\author[1]{Chang Liu}
\author[1,*]{Jianlong Yang}

\affil[1]{School of Biomedical Engineering, Shanghai Jiao Tong University, Shanghai, China}
\affil[2]{The authors contributed equally to this work}

\affil[*]{jyangoptics@gmail.com}

\begin{abstract}
Fiber-form optics extends the high-resolution tomographic imaging capabilities of Optical Coherence Tomography (OCT) to the inside of the human body, i.e., endoscopic OCT. However, it still faces challenges due to the trade-off between probe size, resolution, and Depth Of Focus (DOF). Here we introduce a method for extending the DOF in endoscopic OCT with high uniformity and efficiency. On the basis of multi-level diffractive optics, we leverage the multi-dimensional light-field modulation capabilities of Computer-Generated Holography (CGH), to achieve precise control of the intensity distribution of the off-axis portion of the OCT probe light. Our method eliminates the need for an objective lens, allowing for direct fabrication at the distal facet of a single-mode fiber using femtosecond laser two-photon 3D printing. The superiority of our method has been verified through numerical simulation, beam measurement, and imaging results obtained with our home-built endoscopic OCT system.  
\end{abstract}

\setboolean{displaycopyright}{false} % Do not include copyright or licensing information in submission.

\begin{document}

\maketitle

\begin{figure}
    \centering
    \includegraphics[width=0.95\linewidth]{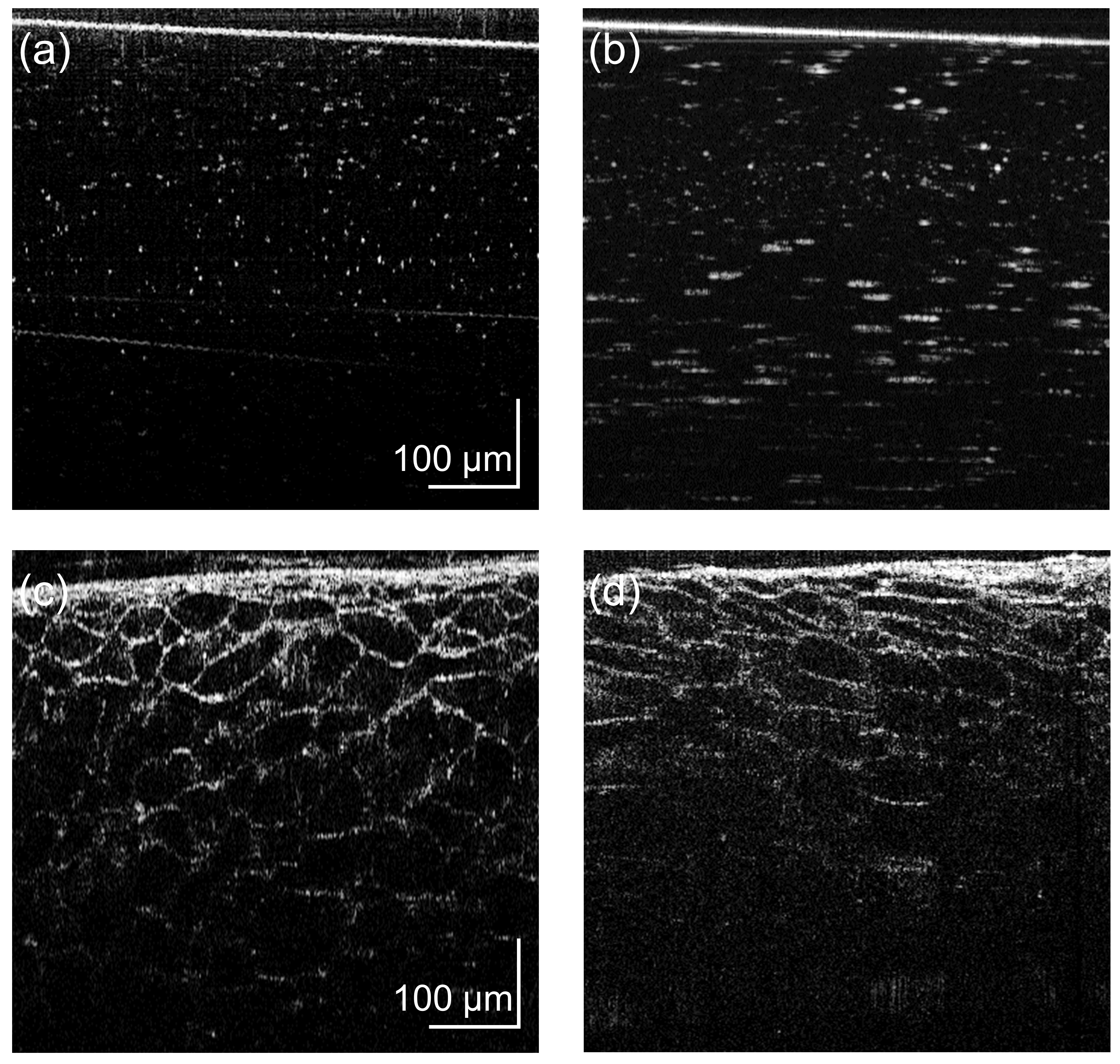}
    \caption{Comparison of the OCT imaging results using the endoscopic probe via our DOF extension method [(a) and (c)] and a free-space Gaussian-beam setup with a similar focusing spot [(b) and (d)]. (a) and (b) are the results of imaging a resolution phantom, which was made of gold nanorods (with dimensions of < 100 nm) and polydimethylsiloxane. (c) and (d) are the result of imaging fresh grape pulp.}
    \label{fig:image}
\end{figure}

Using fiber optics to convert the sample arm into a flexible micro-probe, it is possible to extend the high-resolution tomographic imaging capabilities of Optical Coherence Tomography (OCT) to the inside of the human body, i.e., endoscopic OCT \cite{gora2017endoscopic}. It has been applied in clinical diagnosis and treatment of coronary arteries and intracranial, and plays an important role in assessing the plaque morphology of the inner wall of blood vessels, evaluating the effect of stent implantation, and detecting the stenosis of intracranial arteries \cite{bouma2017intravascular}. On the other hand, OCT in natural cavities (digestive, respiratory, and reproductive tracts, etc.) can be used as an in situ and immediate “optical biopsy”, and is gradually being applied in clinical practice \cite{tsai2014endoscopic,zeng2018ultrahigh,nandy2021diagnostic}. In recent years, endoscopic OCT has also begun to be used to monitor treatment processes such as radiofrequency ablation of the heart and esophagus \cite{huang2023automated,lo2019balloon}, and resection of colorectal cancer \cite{mora2020steerable}, to achieve the integration of precision diagnosis and treatment.\\
\indent However, endoscopic OCT technoloy still suffers from the contradiction between probe size, resolution and imaging range (working distance and Depth Of Focus, DOF). Most existing endoscopic OCT probes use refractive optics such as gradient refractive index (GRIN) lenses/fibers, ball lenses, etc. \cite{gora2017endoscopic}, and their size is usually positively correlated with the focal length. In the pursuit of higher vessel and lumen passage rates, as well as to achieve a higher degree of integration with other imaging and therapeutic instruments, endoscopic OCT probes are moving towards ever finer dimensions \cite{lorenser2012ultrathin,li2020ultrathin,qiu2020uniform}, but this usually results in a reduction of the imaging range. On the other hand, the DOF and working distance can be effectively expanded by changing the refractive index gradient and lens curvature \cite{kang2022pencil,ryu2008lensed}, but at the expense of transverse resolution.\\
\indent Various methods have been proposed to address the DOF issue in endoscopic OCT, including Bessel beams \cite{tan2009fiber}, binary phase masks \cite{lorenser2012ultrathin}, dispersion effects \cite{pahlevaninezhad2018nano}, mode interference \cite{qiu2020uniform}, and dynamic focusing \cite{liao2017endoscopic}. However, these methods are generally limited by lower efficiency and beam quality, less flexibility, and incompatibility with chromatic aberration and astigmatism correction approaches.\\
\indent Recently, Zhao et al. developed a spatially multiplexed phase mask that generates many axially spaced focus points for DOF extension \cite{zhao2022flexible} (hereafter referred to as multi-focal method). This approach allows flexible manipulation of beam length and diameter, but there is still room for improvement in beam uniformity and insertion loss. Besides, it requires the combination of an objective lens and a phase mask, making it inconvenient to apply directly to endoscopic OCT.\\
\indent In this Letter, we introduce a method for high-uniformity, high-efficiency DOF extension in endoscopic OCT. On the basis of multi-level diffractive optics similar to that in \cite{zhao2022flexible}, we leverage the multi-dimensional light-field modulation capabilities of Computer-Generated Holography (CGH) \cite{pi2022review}, to achieve precise control of the intensity distribution of the off-axis portion of the OCT probe light. Using femtosecond laser two-photon 3D printing \cite{gissibl2016two}, we additively fabricate a CGH-optimized phase mask at the distal facet of the single-mode fiber sample arm for endoscopic OCT. The superiority of our method has been verified through numerical simulation and experiments.\\
\begin{figure*}[hhh]
\centering
    \includegraphics[width=\linewidth]{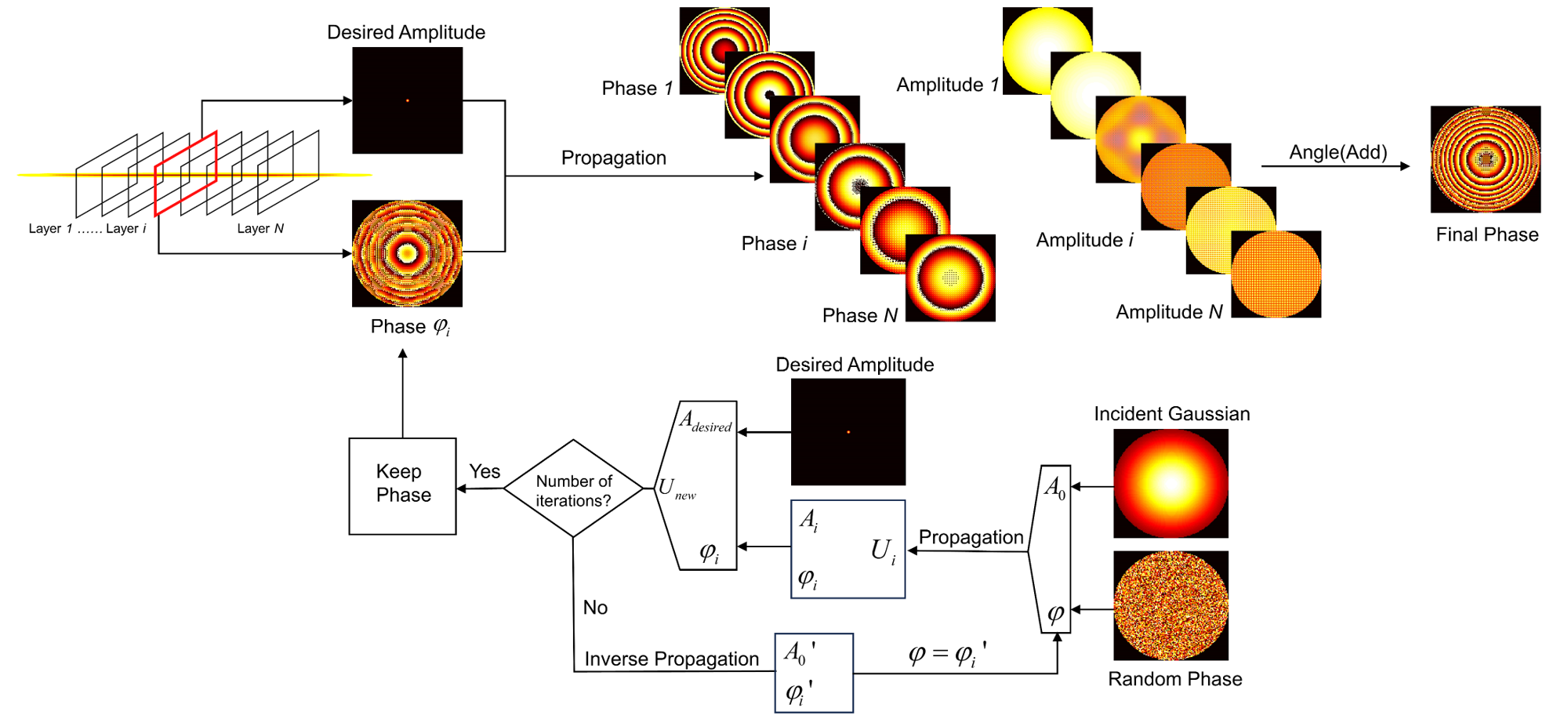}
    \caption{Methodology of producing needle-shaped beams using computer-generated holography.}
    \label{fig:algorithm}
\end{figure*}
\indent Figure~\ref{fig:algorithm} illustrates the methodology of producing needle-shaped beams using CGH.
We adopt the layer-based CGH method \cite{pi2022review} for 3D reconstruction of the target needle beam, i.e., the object is divided into a number of parallel planes, each of which independently serves as a computational unit, and is separately back-propagated to the holographic plane for complex amplitude superposition. We split the three-dimensional light field distribution of the needle-shaped beam into $N$ layers, and the single-layer two-dimensional light field distribution consists of the desired amplitude $A_{desired}$ and phase $\phi_i$, respectively. The desired amplitude is uniformly distributed in a circular region, the diameter of the circle is set according to the demand for the lateral resolution of endoscopic OCT, while the phase needs to be obtained by a phase recovery algorithm under the condition of maintaining the amplitude of the incident light and the desired amplitude constant.\\
\indent Here we adopt the Gerchberg-Saxton (GS) algorithm \cite{Gerchberg1972} for the phase recovery. Its iterative process can be summarized as follows: during the forward and inverse Fourier transform, the outgoing surface amplitude and the observed surface amplitude are continuously replaced with the desired values, while the phase is always kept consistent with the computed results. The process is equivalent to a stochastic gradient descent algorithm with a fixed step size. In this study, we replace the iterative process of forward, inverse Fourier transform in the GS algorithm with the corresponding scalar diffraction propagation process. Using the phase recovery algorithm to find the phase distribution of a single-layer 2D light field has two advantages: 1) the searched phase can match the amplitude of the incident light (e.g., Gaussian distribution), and 2) higher focusing efficiency is expected to be achieved by constructing a desired energy distribution.\\
\indent The input to the GS algorithm is a Gaussian beam with random phase distribution:
\begin{equation}
    U(x,y,z=0)=A\exp [-({{x}^{2}}+{{y}^{2}})/{{r_0}^{2}}]\exp[j\varphi(x,y,z=0)],
    \label{gb}
\end{equation}
where $A$ is the amplitude of the beam, $(x,y)$ is the coordinates of the plane perpendicular to the direction of beam propagation. $z$ is the coordinates along the direction of beam propagation. $r_0$ is the $1/{{e}^{2}}$ radius of the input Gaussian beam. $\varphi(x,y,z=0)$ is the random phase. Then the complex amplitude distribution on the ${{z}_{i}}$ plane (a single layer) can be expressed by Fresnel diffraction \cite{goodman2005introduction}:\\
\begin{equation}
U(x,y,z_i) = \mathcal{F}^{-1} \left\{ \mathcal{F} \left\{ U(x,y,0) \right\} H(f_X, f_Y, z_i) \right\},
\end{equation}
where $H$ is the Fresnel transfer function. $\mathcal{F} $ and $\mathcal{F}^{-1}$ represents forward and inverse Fourier transforms, respectively. $({{f}_{X}},{{f}_{Y}})$ is the counterparts of $(x,y)$ in frequency space. $H$ can further be written as:
\begin{equation}
    H({{f}_{X}},{{f}_{Y}},{{z}_{i}})={{e}^{jk{{z}_{i}}}}\exp [-j\pi \lambda {{z}_{i}}({{f}_{X}}^{2}+{{f}_{Y}}^{2})],
\end{equation}
where $\lambda$ is the wavelength of the incident beam, $k=\frac{2\pi }{\lambda}$ is the corresponding wavenumber. \\
\indent To achieve the targeted needle-shaped beam, we replace the amplitude of the ${{z}_{i}}$ plane with the desired amplitude ${{A}_{desired}}$ and keeping the phase ${{\phi }_{i}}$ unchanged. The new complex amplitude distribution can be written as:
\begin{equation}
    {{U}_{new}}(x,y,{{z}_{i}})={{A}_{desired}}\exp [(j\cdot angle(U(x,y,{{z}_{i}})))].
\end{equation}
Then the backpropagation from the ${{z}_{i}}$ plane to the incident plane $z=0$ can be calculated with:
\begin{equation}
U(x,y,0) = \mathcal{F}^{-1} \left\{ \mathcal{F} \left\{ U_{new}(x,y,z_i) \right\} \cdot H(f_X, f_Y, -z_i) \right\}.
\end{equation}
\indent We replace the amplitude of $U(x,y,0)$ with the amplitude of the input Gaussian beam in Eq.~\ref{gb}. While the phase $\varphi (x,y,0)=angle[U(x,y,0)]$ is kept unchanged to form a new complex amplitude distribution for the next iteration:
\begin{equation}
{{U}_{new}}(x,y,0)=A\exp [-({{x}^{2}}+{{y}^{2}})/{{r_0}^{2}}]\exp (j\varphi (x,y,0)).    
\end{equation}
\indent The iterative process is executed repeatedly, converging towards the solution as we approach the predetermined threshold of iterations. Then we preserve the ${{z}_{i}}$ plane phase ${{\phi }_{i}}$ and combine with the desired amplitude ${{A}_{desired}}$ to achieve the construction of the single-layer light field distribution.\\
\indent Finally, we summarize the complex amplitudes at $z=0$ of all $N$ layers, and use its phase as the phase distribution of multi-level diffractive optical element (DOE):
\begin{equation}
\phi_{final} = angle\left\{\sum^N[U_{z_i}(x,y,0)]\right\}.
   \label{sum} 
\end{equation}
For the numerical simulation and manufacturing of the multi-level DOE, we quantize $\phi_{final}$ into 16 height levels (same as the level number in \cite{zhao2022flexible} for a fair comparison). So each level corresponds to a phase change of 0.125$\pi$. The height unit is 95 nm based on the central wavelength (840 nm) of our home-built endoscopic OCT system \cite{zhang2024cross}. The transverse plane is divided into pixels, which are sized by $1\times 1$ $\mu$m$^2$.\\
\begin{figure}
    \centering
    \includegraphics[width=0.95\linewidth]{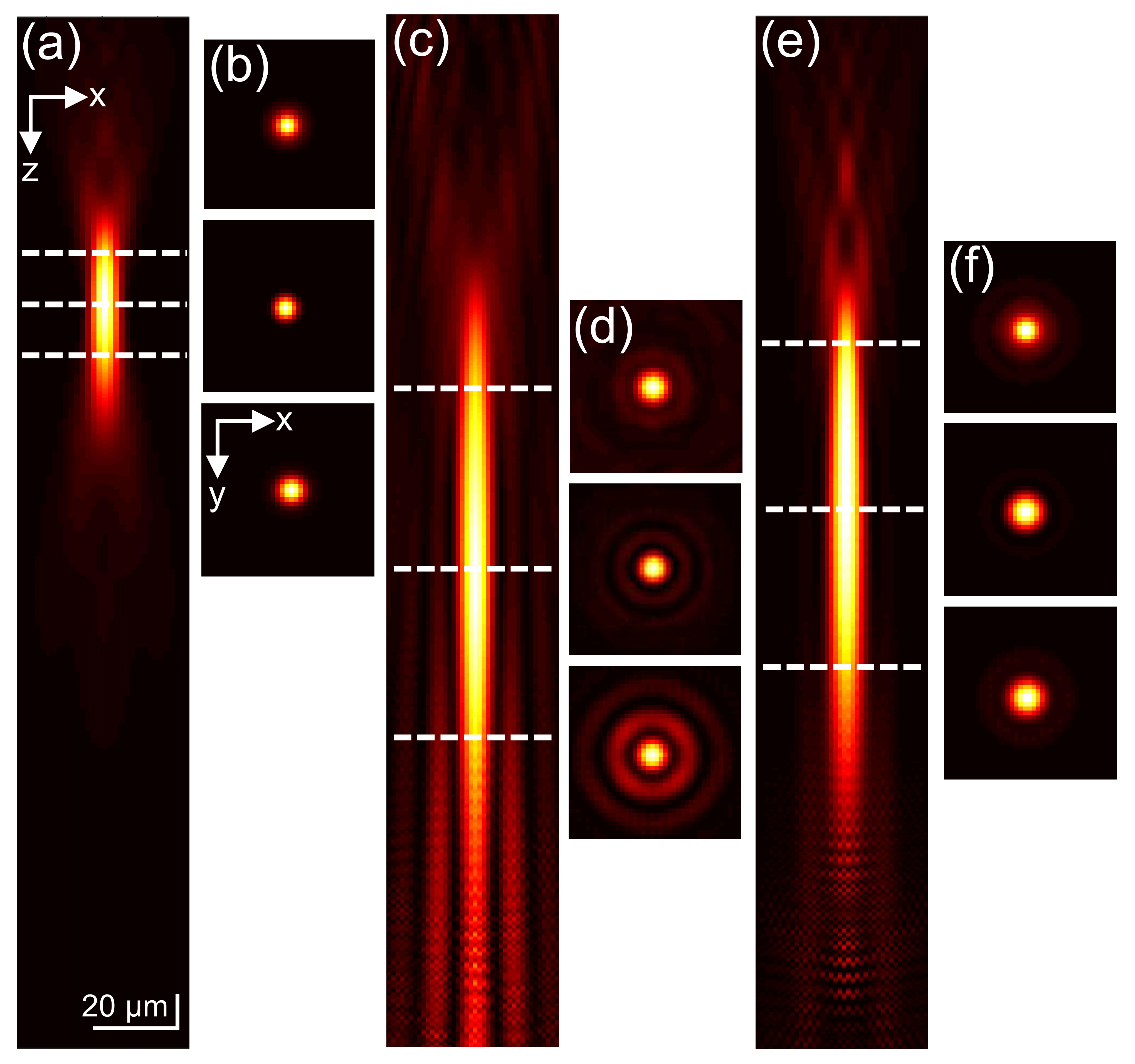}
    \caption{Comparison of numerical simulation results for generating a focused beam using different methods. (a) and (b) are the results of Gaussian beam. (c) and (d) are the results of the needle-shaped beam via the multi-focal method in \cite{zhao2022flexible}. (e) and (f) are the results of the needle-shaped beam via the CGH method proposed in this work.}
    \label{fig:simu}
\end{figure}
\indent We first perform numerical simulation to validate the effectiveness of our method. In the simulation, we set the target resolution to 5 $\mu$m, i.e., the Full Width at Half Maximum (FWHM) of the beam. We also assume an aperture of 125 $\mu$m, which is the diameter of the inner cladding of single-mode fibers. The baselines include Gaussian beams and the needle-shaped beams via the multi-focal method \cite{zhao2022flexible}. The Gaussian beams are obtained by passing the beam through a convex lens phase with a focal length of 500 $\mu$m and propagating it using Fresnel diffraction. We follow the description in \cite{zhao2022flexible} to realize the multi-focal method. In the target focal region (500 to 700 $\mu$m behind the phase mask), we establish 9 equidistant focal points along the axial ($z$) direction, and the phase adjuster is set to 0.028$\pi$ to generate needle-shaped beams. In our method, we divided the range of 500 to 700 $\mu$m behind the phase mask into 80 planes, and the expectation for each plane was set to an ideal spot with a FWHM of 5 $\mu$m.\\
\indent Figure~\ref{fig:simu} demonstrates the comparison of numerical simulation results for generating a focused beam using different methods. (a) and (b) are the results of Gaussian beam. (c) and (d) are the results of the needle-shaped beam via the multi-focal method in \cite{zhao2022flexible}. (e) and (f) are the results of the needle-shaped beam via the CGH method proposed in this work. (a), (c), and (e) are the cross-sections of the beams at the $x-z$ plane (the results are similar at the $y-z$ plane, because the generated beam is rotationally symmetric in the z-axis). (b), (d), and (f) are their corresponding cross-sections at the $x-y$ plane, at the positions of the dashed lines in (a), (c), and (e). As shown in the figure, the needle-shaped beams generated from the multi-focal method \cite{zhao2022flexible} and our method have much longer focal regions than that of the Gaussian beam. Compared to the former, our method produces beams with less diffracted side lobes and more uniform axial and transverse distributions.\\
\begin{figure}
    \centering
    \includegraphics[width=\linewidth]{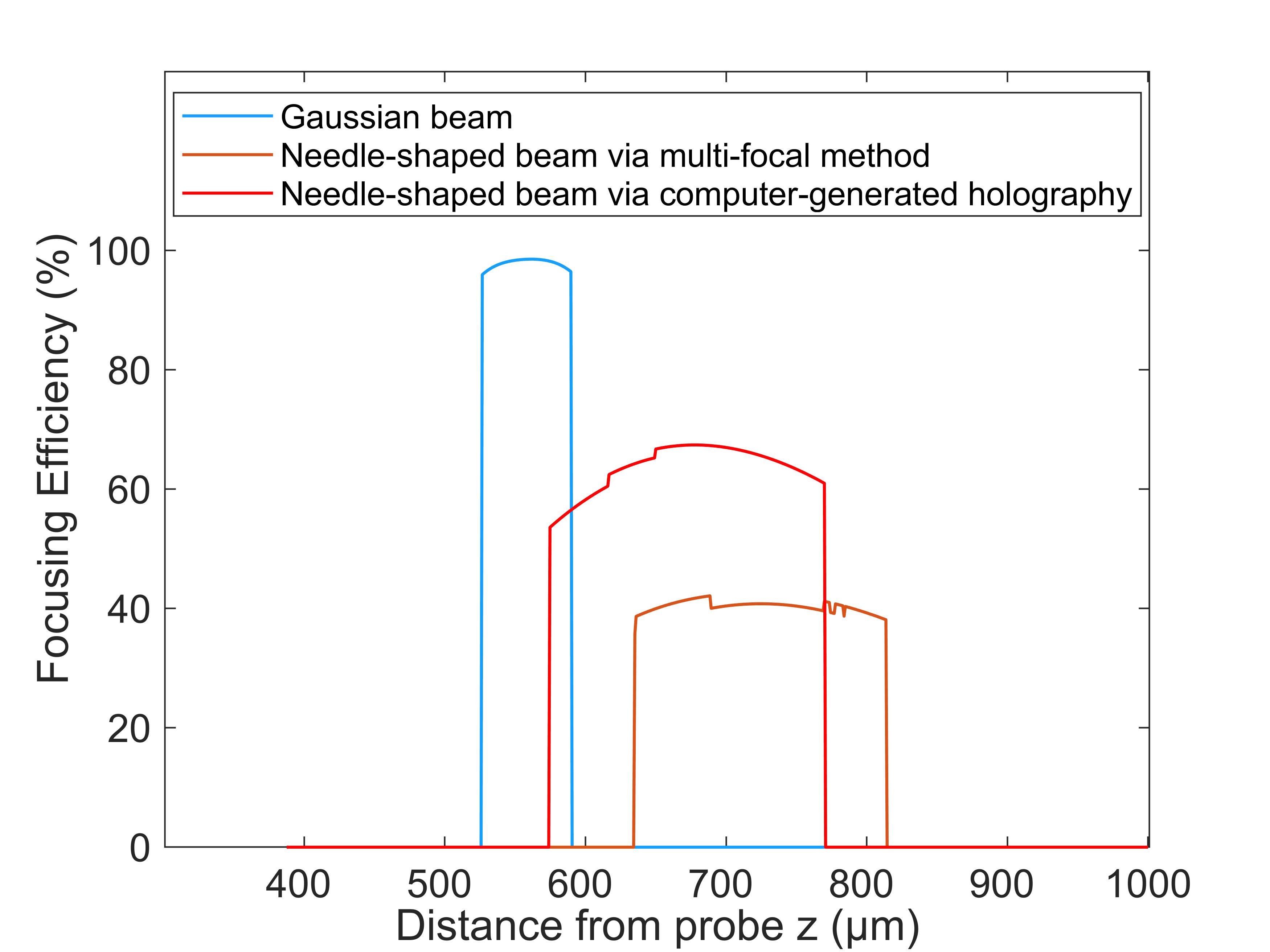}
    \caption{Comparison of focusing efficiency for the generated focused beams using different methods.}
    \label{fig:efficiency}
\end{figure}
\indent We further compare their performance in focusing efficiency as plotted in Fig.~\ref{fig:efficiency}. The focusing efficiency is defined as the percentage of the total incident Gaussian energy in a circle of 3 times the FWHM diameter in each $x-y$ plane distributed along the axial direction ($z$) \cite{arbabi2015subwavelength}. Among them, the Gaussian beam has the highest focusing efficiency to about 98\%, but its short DOF limits its imaging performance in endoscopic OCT, as shown in Fig.~\ref{fig:image} (a) and (c). The diffractive optics methods, including the multi-focal \cite{zhao2022flexible} and our methods, are relatively inefficient. This is due to the fact that we only used a DOE with 16 height levels. In the future, with the improvement of processing accuracy, the focusing efficiency can be further improved by using DOEs with more height levels \cite{khonina2024exploring}. Even so, our method still shows a substantial improvement in focusing efficiency compared to the multifocal method \cite{zhao2022flexible}, with the maximum focusing efficiency increasing from about 42\% to 67\%.\\
\begin{figure}
    \centering
    \includegraphics[width=\linewidth]{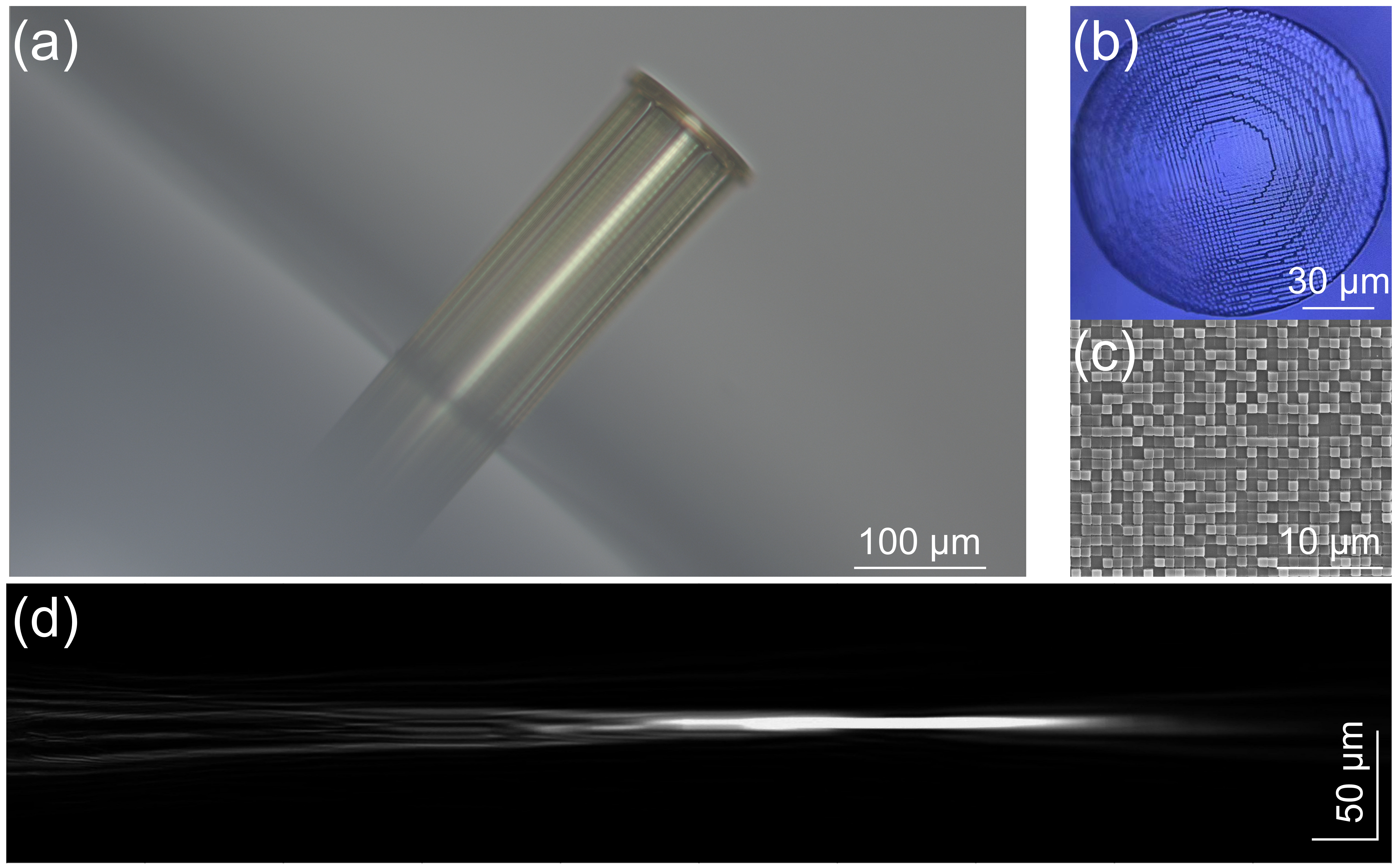}
    \caption{Details of the endoscopic OCT probe using our DOF extension method. (a) and (b) are its microscopic photos from side and top views, respectively. (c) is the photo of the DOE's surface using a scanning electron microscope. (d) is the measured beam profile of the probe.}
    \label{fig:photo}
\end{figure}
\indent Using a commercial femtosecond laser two-photon 3D printer, we fabricated the DOEs designed via our method at the distal end of single-mode fibers. We first printed a section of a solid cylinder about 690 $\mu$m long on the end facet of a single-mode fiber, which expands the beam diameter to about 100 $\mu$m. Then the designed DOE was printed on top of the cylinder. The photoresist material we used for the 3D printing has a refractive index of 1.56. Figure~\ref{fig:photo} gives the details of the endoscopic OCT probe using our DOF extension method. (a) and (b) are its microscopic photos from side and top views, respectively. (c) is the photo of the DOE's surface using a scanning electron microscope. (d) is the measured beam profile of the probe. We used the light source of our endoscopic OCT system \cite{zhang2024cross}, which has a central wavelength of 840 nm and a bandwidth of about 50 nm. The beam measurement was performed by an imaging system consisting of a 20$\times$ objective lens, a lens with a focal length of 75 mm, and a camera. We view the light exiting the end face of the optical fiber as the object, which forms a conjugate relationship with the image received by the camera. Keeping the camera stationary, we gradually moved the translation stage of the fiber, to record the transverse light-field distribution surface-by-surface and reconstruct the profile. The measured beam has a FWHM spot size of 5 $\mu$m and a DOF of 224 $\mu$m. As a comparison, the Gaussian beam with the same spot size has a DOF of 67 $\mu$m. We also measured the insertion loss of our probe with an optical power meter. For the input power to the fiber of 17 mW, the output power from the DOE is 12.4 mW, which leads to an insertion loss of 1.367 dB.\\
\indent Figure~\ref{fig:image} shows the comparison of the OCT imaging results using the endoscopic probe via our DOF extension method [(a) and (c)] and a free-space Gaussian-beam setup with a similar focusing spot [(b) and (d)]. (a) and (b) are the results of imaging a resolution phantom, which was made of gold nanorods (with dimensions of < 100 nm) and polydimethylsiloxane. (c) and (d) are the result of imaging fresh orange pulp. It can be seen that our method is able to maintain a high resolution over the entire imaging range, whereas the Gaussian beam with a similar resolution would show a significant loss of resolution and signal strength due to defocusing.\\
\indent In summary, we have developed a CGH-based method for the DOF extension in endoscopic OCT, which has achieved higher uniformity and efficiency than the existing ones. Using femtosecond laser two-photon 3D printed, we have fabricated the proposed probe directly on the distal end facet of a single-mode fiber. The superiority of our method has been verified through numerical simulation, beam measurement, and imaging results obtained with our home-built endoscopic OCT system. Further improvements on this work could be increasing the height levels of DOEs, optimizing the CGH algorithms for DOF extension, and considering the astigmatism and aberration in the design.
\begin{backmatter}
\bmsection{Funding} National Natural Science Foundation of China (62105198).
\bmsection{Disclosures} The authors declare no conflicts of interest.
\bmsection{Data availability}
Data underlying the results presented in this paper are not publicly available at this time but may be obtained from the authors upon reasonable request.
\end{backmatter}
% Bibliography
\bibliography{sample}

% Full bibliography added automatically for Optics Letters submissions; the following line will simply be ignored if submitting to other journals.
% Note that this extra page will not count against page length
\bibliographyfullrefs{sample}

%Manual citation list
%\begin{thebibliography}{1}
%\bibitem{Zhang:14}
%Y.~Zhang, S.~Qiao, L.~Sun, Q.~W. Shi, W.~Huang, %L.~Li, and Z.~Yang,
 % \enquote{Photoinduced active terahertz metamaterials with nanostructured
  %vanadium dioxide film deposited by sol-gel method,} Opt. Express \textbf{22},
  %11070--11078 (2014).
%\end{thebibliography}

% Please include bios and photos of all authors for aop articles
\ifthenelse{\equal{\journalref}{aop}}{%
\section*{Author Biographies}
\begingroup
\setlength\intextsep{0pt}
\begin{minipage}[t][6.3cm][t]{1.0\textwidth} % Adjust height [6.3cm] as required for separation of bio photos.
  \begin{wrapfigure}{L}{0.25\textwidth}
    \includegraphics[width=0.25\textwidth]{john_smith.eps}
  \end{wrapfigure}
  \noindent
  {\bfseries John Smith} received his BSc (Mathematics) in 2000 from The University of Maryland. His research interests include lasers and optics.
\end{minipage}
\begin{minipage}{1.0\textwidth}
  \begin{wrapfigure}{L}{0.25\textwidth}
    \includegraphics[width=0.25\textwidth]{alice_smith.eps}
  \end{wrapfigure}
  \noindent
  {\bfseries Alice Smith} also received her BSc (Mathematics) in 2000 from The University of Maryland. Her research interests also include lasers and optics.
\end{minipage}
\endgroup
}{}

\end{document}